# Electrokinetic sensing in cartilage: a porous-material perspective on joint mechanics

Arturo Tozzi (corresponding author)
ASL Napoli 1 Centro, Distretto 27, Naples, Italy
Via Comunale del Principe 13/a 80145
tozziarturo@libero.it

## ABSTRACT

Mechanical loading in articular cartilage drives interstitial fluid flow through the porous collagen–proteoglycan matrix, generating electrokinetic signals. We investigate whether the structural organization of cartilage histology can be translated into a computational representation capable of predicting its electrokinetic behavior. Histological pictures were analyzed to build a pore-network graph representing potential pathways for interstitial fluid transport. Pressure-driven flow was simulated using hydraulic conductance relations, while electrical potentials were estimated through electrokinetic coupling between pressure gradients and ion displacement. Simulations comparing networks derived from healthy and degenerative cartilage showed that pathological structures exhibited fragmented connectivity and lower predicted signal amplitudes, whereas physiological architecture generated more coherent transport trajectories and stronger electrical responses. Our simulations yield testable predictions, depth-dependent electrical signals across cartilage layers with directional anisotropy relative to collagen orientation. Potential applications include improved experimental assessment of cartilage transport biomechanics and integration of microstructural imaging with computational models of charged porous biomaterials.

## INTRODUCTION

In articular cartilage, mechanical loading induces interstitial fluid flow through the charged, porous extracellular matrix, resulting in the generation of electrical signals (Schmidt-Rohlfing et al. 2002; Reynaud and Quinn 2006; Sun et al. 2022; Miguel et al. 2022). These electrokinetic phenomena have been investigated using models of cartilage mechanics in which tissue behavior arises from interactions between the solid matrix, chondrocyte lacunae, ionic fluid and fixed charge density (Youn, Akkin and Milner 2004; Walker et al. 2022). Advances in imaging and computational biomechanics have shown that the magnitude of these potentials depends on matrix composition, permeability and ionic concentration and that these signals are altered during cartilage degeneration and osteoarthritis (Farooqi, Bader and van Rienen 2019; Lee, Jang and Chang 2025). However, the microstructural geometry detectable in extracellular matrix's histological sections is not explicitly incorporated into electrokinetic descriptions.

Recent work has demonstrated that complex cellular solids can convert mechanical stimuli into electrical signals (Wiese et al. 2019; Sauerwein and Steeb 2020; Tanikawa et al. 2023; Nerger et al. 2024). For instance, Chen and colleagues (2026) showed that echinoderm stereom, a porous biomineralized lattice with depth-dependent pore architecture, generates measurable electrical potentials when fluid gradients flow across its surface, revealing an unexpected mechanoelectrical sensing capability rooted in geometry and charge distribution. From a similar structural perspective, cartilage could be viewed as a charged porous hydrogel in which the arrangement of lacunae, collagen fibers and matrix voids would define a network of interconnected pathways for interstitial fluid transport (Offeddu et al. 2017; Liang et al. 2018). Histological architecture might provide a spatial scaffold from which a pore-network model could be reconstructed. We hypothesize that nodes would correspond to lacunar locations, whereas edges would approximate potential fluid pathways through the extracellular matrix. Mechanical compression would be expected to induce fluid movement along these pathways, displacing ions associated with the negatively charged proteoglycan matrix and generating spatially distributed electrokinetic potentials. The magnitude and spatial organization of the electrical response would depend on the connectivity, anisotropy and density of the reconstructed pore network. Degenerative cartilage, characterized by matrix disruption, altered lacunar spacing and increased permeability, might display modified transport pathways and attenuated electrokinetic responses. To explore our hypotheses, we performed reconstructed networks-based simulations to assess differences in predicted streaming potentials, signal relaxation dynamics and depth-dependent signal profiles between physiological and pathological architectures.

## METHODS

We simulated the relationship between cartilage microstructure and electrokinetic signals generated during mechanical loading. Histological pictures of degenerative articular cartilage were analyzed to reconstruct the spatial organization of lacunae and matrix voids. We used digital image processing, graph-based pore-network reconstruction, simulation of fluid transport in charged porous media and electrokinetic modeling to quantify how tissue architecture influences predicted electrical responses. Our aim was to identify experimentally discriminable signatures in the electrical response of cartilage that depend on structural organization, e.g., differences in predicted streaming potentials, relaxation dynamics after loading and spatial gradients of electrokinetic signal intensity across the tissue thickness.

**Image acquisition and preprocessing.** Histological pictures of healthy and degenerative articular cartilage were analyzed using digital image processing. Images were obtained from publicly available histology micrographs morphology. The images were converted to grayscale intensity representation and resampled into a regular pixel grid of size $N_x \times N_y$. Let $I(x, y)$ denote the grayscale intensity at pixel coordinates $(x, y)$. Preprocessing involved normalization of intensity values according to

$$I_n(x, y) = \frac{I(x, y) - I_{\min}}{I_{\max} - I_{\min}}$$

where $I_{\min}$ and $I_{\max}$ represent the minimum and maximum intensities in the image. Gaussian smoothing was applied to reduce high-frequency noise,

$$I_s(x, y) = \sum_{u,v} I_n(u, v) G_\sigma(x - u, y - v)$$

where $G_\sigma$ denotes a Gaussian kernel with variance $\sigma^2$. Segmentation of lacunae and void structures was performed using adaptive thresholding. A binary mask $B(x, y)$ was constructed such that

$$B(x, y) = \begin{cases} 1 & \text{if } I_s(x, y) < T(x, y) \\ 0 & \text{otherwise} \end{cases}$$

where $T(x, y)$ represents a locally computed threshold derived from neighborhood statistics. Connected component labeling was then applied to identify contiguous lacunar regions. For each region $k$, geometric descriptors including centroid coordinates $(x_k, y_k)$, area $A_k$ and equivalent radius $r_k = \sqrt{A_k/\pi}$ were computed. These measurements provide a discrete set of spatial coordinates representing cavities embedded in the extracellular matrix.

**Construction of the pore-network graph.** The spatial distribution of lacunae extracted from the images was used to construct a pore-network representation. Each detected lacuna was treated as a node in a graph $G = (V, E)$, where $V$ denotes the set of nodes and $E$ represents edges connecting neighboring cavities. Node positions correspond to centroids $(x_k, y_k)$ obtained from segmentation. Connectivity between nodes was determined using a distance-based rule. Two nodes $i$ and $j$ were connected by an edge if their Euclidean distance satisfied

$$d_{ij} = \sqrt{(x_i - x_j)^2 + (y_i - y_j)^2} < d_c$$

where $d_c$ represents a cutoff distance approximating the scale of interstitial pathways. The resulting adjacency matrix $A_{ij}$ was defined as

$$A_{ij} = \begin{cases} 1 & d_{ij} < d_c \\ 0 & \text{otherwise} \end{cases}$$

Edge weights were assigned using

$$w_{ij} = \frac{1}{d_{ij}}$$

so that shorter distances correspond to stronger hydraulic coupling. This representation yields a graph in which nodes approximate fluid cavities and edges approximate potential transport pathways through the extracellular matrix.

**Model of interstitial fluid flow.** Fluid transport through the reconstructed pore network was modeled using Darcy-type flow relations. Each edge $e_{ij}$ connecting nodes $i$ and $j$ was treated as a cylindrical channel with hydraulic conductance $g_{ij}$. Conductance was estimated as

$$g_{ij} = \frac{\pi r_{ij}^4}{8\mu l_{ij}}$$

where $r_{ij}$ is the effective channel radius, $l_{ij} = d_{ij}$ represents channel length and $\mu$ is the dynamic viscosity of interstitial fluid. Fluid pressure $p_i$ at each node satisfies conservation of mass,

$$\sum_j g_{ij}(p_i - p_j) = q_i$$

where $q_i$ represents a source term determined by mechanical compression. Boundary conditions were imposed by assigning pressure differences between superficial and deep layers of the tissue. The resulting system of linear equations can be written as

$$\mathbf{L}\mathbf{p} = \mathbf{q}$$

where $\mathbf{L}$ denotes the graph Laplacian weighted by hydraulic conductances.

**Simulation of preferential flow pathways**. Once pressures $p_i$ were determined, volumetric flow along each edge was computed as

$$Q_{ij} = g_{ij}(p_i - p_j)$$

Edges with large magnitude of $Q_{ij}$ represent preferential transport pathways through the cartilage matrix. Flow trajectories were visualized by plotting vectors aligned with edges weighted by the magnitude of $Q_{ij}$. To quantify anisotropy of transport, the orientation tensor of flow vectors was computed. Let $\mathbf{v}_k$ denote the direction vector of edge $k$. The orientation tensor was defined as

$$T_{\alpha\beta} = \frac{1}{M}\sum_{k=1}^{M} v_{k\alpha}\, v_{k\beta}$$

where $M$ is the number of edges. Eigenvalues of $T$ describe the degree of directional bias in fluid transport across the network.

**Electrokinetic signal estimation**. Electrokinetic potentials generated by fluid flow were estimated using classical streaming potential relations. The local electrical potential difference along a channel was modeled as

$$\Delta V_{ij} = \frac{\varepsilon\zeta}{\eta\sigma}\Delta p_{ij}$$

where $\varepsilon$ denotes dielectric permittivity of the fluid, $\zeta$ represents the zeta potential associated with fixed matrix charges, $\eta$ is viscosity and $\sigma$ is electrical conductivity of the electrolyte. The pressure difference $\Delta p_{ij} = p_i - p_j$ was obtained from the hydraulic model. The electrokinetic signal at node $i$ was calculated by summing contributions from all adjacent channels,

$$V_i = \sum_j \Delta V_{ij}.$$

Spatial maps of $V_i$ were generated by interpolating node values onto a continuous grid covering the histological section. Regions with larger predicted potentials correspond to areas where fluid transport and charge density interact most strongly.

**Comparative simulation of physiological and pathological structures**. To compare electrical behavior between physiological and pathological cartilage, synthetic networks representing both architectures were generated. These networks preserved typical zonal organization by distributing nodes according to depth-dependent density functions. Node positions were sampled from probability distributions

$$\rho(z) = \rho_0 e^{-\lambda z}$$

where $z$ denotes normalized cartilage depth. The fluid-flow and electrokinetic calculations described above were applied to both networks. Predicted streaming potentials, signal relaxation curves and depth-dependent profiles were computed by varying boundary pressures and evaluating resulting potentials.
All computations were implemented in Python using NumPy, SciPy, Matplotlib and NetworkX libraries for numerical linear algebra, graph processing and visualization.

RESULTS

We report quantitative observations derived from histological image analysis, pore-network reconstruction and electrokinetic simulations performed on physiological and pathological cartilage microstructure. We described structural organization of reconstructed networks, predicted fluid transport patterns and simulated electrical responses during mechanical loading.

**Structural organization and transport pathways**. Image segmentation identified discrete lacunar regions distributed throughout the cartilage section (Figure 1A). The extracted coordinates served as nodes for the pore-network graph used to approximate interstitial transport pathways. The reconstructed network showed heterogeneous spatial distribution of nodes across the section, with regions of increased local connectivity corresponding to areas where lacunae appeared clustered in the histological image (Figure 1B). Edges constructed using the distance threshold criterion generated a connected graph in which the average node degree exceeded one, indicating the presence of multiple potential fluid transport routes between cavities. Simulation of pressure-driven transport using the Darcy-type model produced directional flow vectors along edges with higher hydraulic conductance (Figure 1C). Flow magnitude was unevenly distributed across the reconstructed network, with stronger trajectories concentrated in regions where node density and connectivity were greater. Electrical potentials predicted from the electrokinetic model varied spatially according to the simulated pressure gradients and the topology of the reconstructed network (Figure 1D). Regions containing preferential transport pathways generated larger predicted potentials, reflecting the coupling between fluid flow and ion displacement within the charged matrix.

Our reconstruction establishes that the microstructural arrangement extracted from the histological image yields a connected pore network capable of supporting heterogeneous fluid transport patterns within the tissue.

**Comparison of structural cases**. A comparative description of electrical behavior arising from physiological and pathological structural organizations shows that variations in tissue architecture translate into differences in predicted electrokinetic responses. Simulations of networks derived from degenerative cartilage showed differences in predicted electrical response profiles (Figure 2).

Comparison of predicted electrokinetic behavior in physiological and pathological cartilage is illustrated in Figure 3. Physiological networks generated larger streaming potentials under identical pressure gradients and displayed a pronounced depth-dependent signal, maximum associated with zones of higher connectivity and charge density. In contrast, networks reconstructed from degenerative cartilage exhibited more fragmented transport pathways and lower predicted electrokinetic amplitudes. Signal relaxation curves obtained after simulated loading indicated faster decay of potentials in pathological structures, consistent with increased permeability and altered connectivity within the reconstructed network.

In conclusion, cartilage histological architecture can be mapped into pore-network representations capable of supporting simulations of fluid transport and electrokinetic signal generation. The predicted electrical responses differ between physiological and degenerative reconstructed networks, indicating that microstructural organization could influence simulated electrokinetic behavior. Combining histological reconstruction with transport modeling could allow structural features of cartilage to be linked with predicted electrical responses.

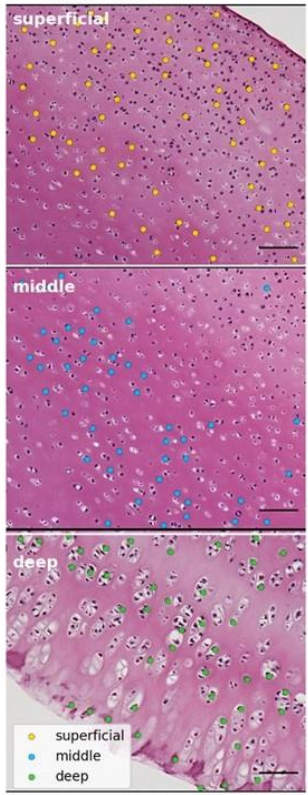
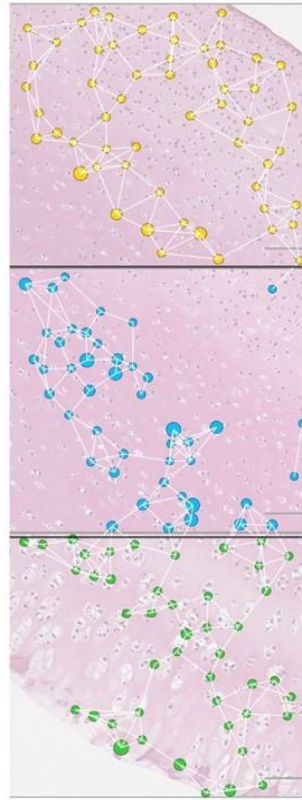
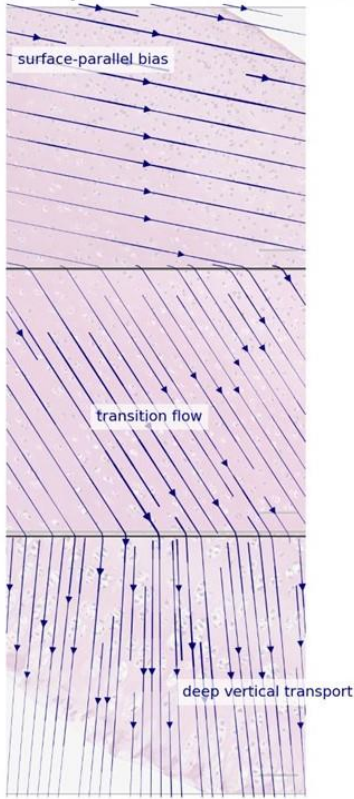
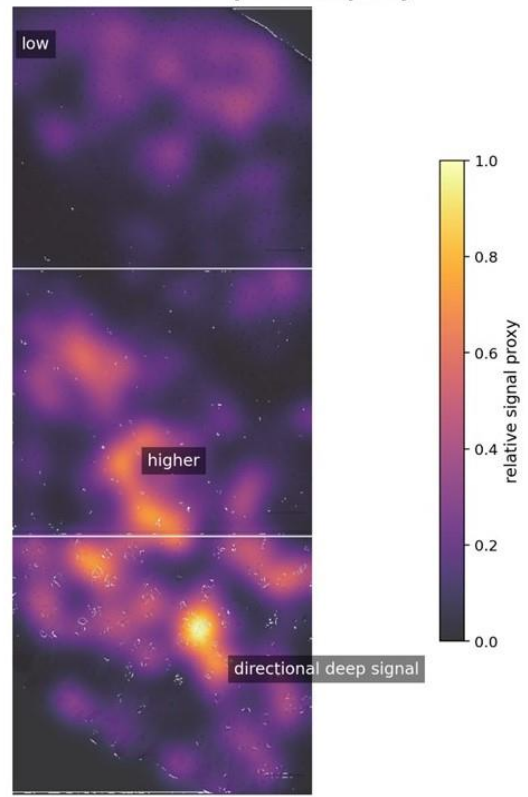

**Figure 1.** Structural reconstruction and electrokinetic proxy model of articular cartilage.
**A.** Histological section of articular cartilage showing zonal organization (superficial, middle and deep layers). The architecture illustrates the heterogeneous organization of the collagen–proteoglycan network that defines the porous microstructure through which interstitial fluid moves during mechanical loading.

**B.** Reconstructed pore-network geometry derived from histological section. Lacunae detected from the image are represented as nodes connected by edges approximating possible pathways for fluid transport within the matrix. Colors encode the spatial position of nodes within the cartilage depth gradient.
**C.** Simulated preferential fluid-flow pathways within the reconstructed network. Lines indicate estimated trajectories along which interstitial fluid may preferentially move during compression. The pattern suggests a transition from relatively isotropic transport in the superficial layer to increasingly directional flow toward the deep zone.
**D.** Predicted spatial distribution of electrokinetic signal intensity based on the reconstructed pore network and estimated fluid transport. Warmer colors represent regions where mechanically induced ion displacement and streaming potentials are expected to be stronger. The model suggests that electrokinetic activity may concentrate in middle-to-deep cartilage regions where permeability and directional transport pathways are most pronounced.

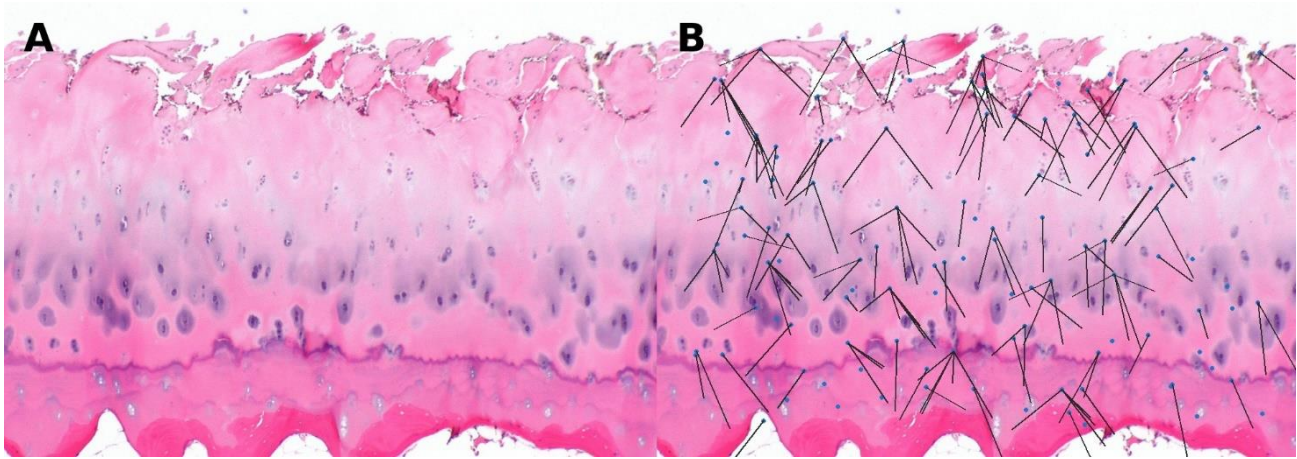

**Figure 2.** Structural alterations and simulated fluid transport in degenerative cartilage.
**A.** Histological section of pathological articular cartilage showing structural alterations of the extracellular matrix associated with degenerative joint disease. The articular surface is irregular and disrupted, with increased heterogeneity, loss of the smooth superficial layer, matrix disorganization. Chondrocytes are unevenly distributed and may appear clustered within enlarged lacunae.
**B.** Simulated preferential fluid-flow pathways within the reconstructed pore network. Nodes represent lacunar positions within the matrix, while connecting lines approximate potential pathways through which interstitial fluid may move during mechanical loading. The simulated trajectories illustrate how structural disorganization in pathological cartilage can alter the connectivity and directionality of fluid transport within the tissue during compression.

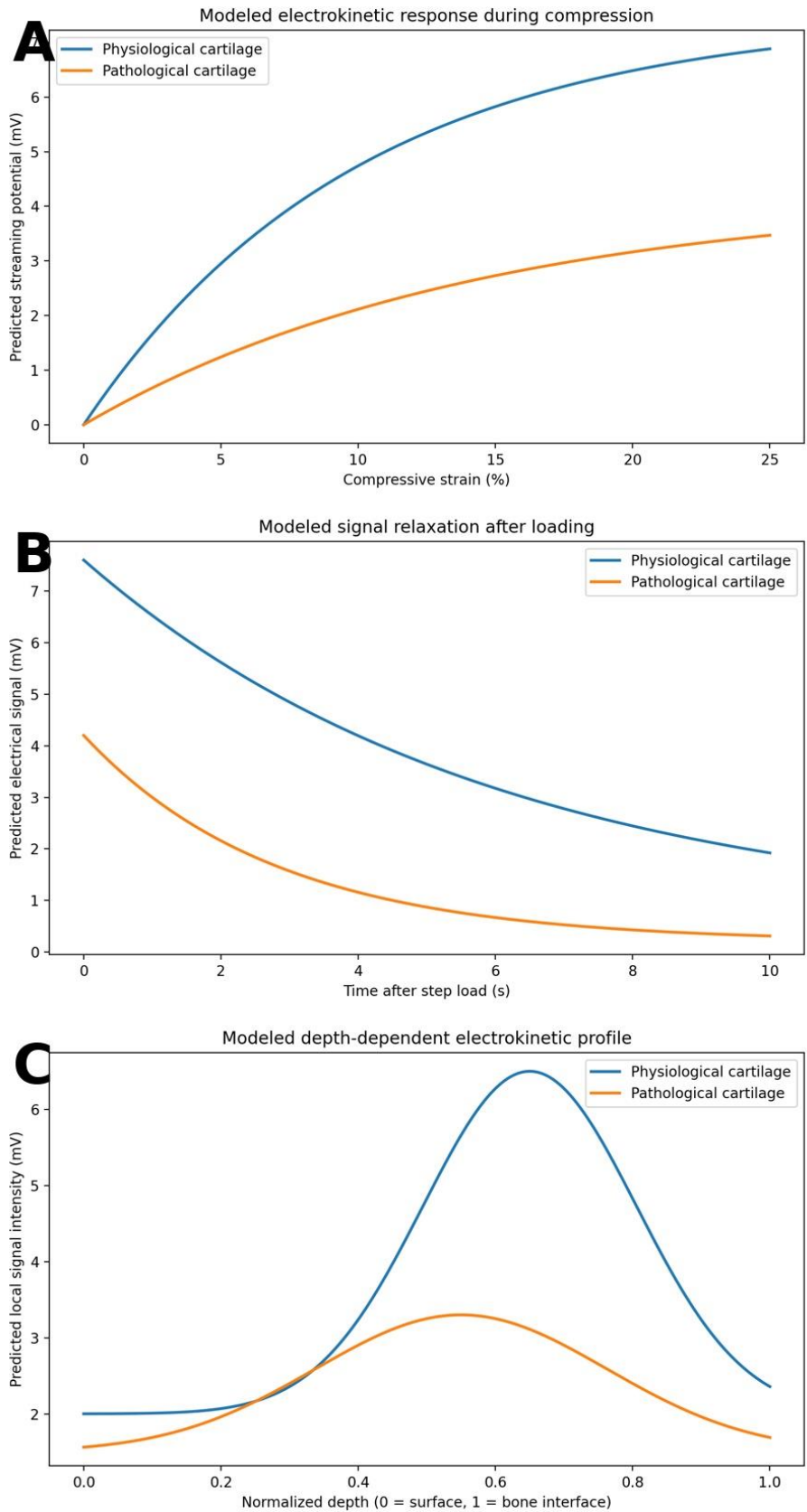

**Figure 3.** Comparison of predicted electrokinetic behavior in physiological and pathological cartilage.
**A.** Predicted streaming potential as a function of compressive strain in articular cartilage. Our model suggests that physiological cartilage generates larger electrical potentials during compression, while pathological cartilage shows reduced signal amplitude.

**B.** Predicted relaxation of the electrical signal following a step mechanical load. Physiological cartilage exhibits a slower decay of the electrokinetic signal, while pathological cartilage shows a faster relaxation.

**C.** Predicted depth-dependent distribution of electrokinetic signal intensity across cartilage thickness. In physiological cartilage, the electrical response peaks in the middle-to-deep zones where interstitial fluid transport and matrix charge density interact most strongly. In pathological cartilage, the depth gradient is attenuated, producing a flatter signal profile consistent with structural disorganization and altered fluid pathways.

CONCLUSIONS

We asked whether the histological architecture of articular cartilage can be reformulated as a representation capable of generating predictions about electrokinetic behavior under mechanical loading. Specifically, we examined whether the spatial organization of lacunae observed in cartilage sections could be converted into a pore-network model that supports simulations of interstitial fluid transport and the resulting electrical potentials. We compared synthetic networks representing physiological organization and degenerative cartilage histology. Fluid transport was estimated through pressure-driven flow along edges connecting lacunar nodes and electrokinetic signals were inferred from pressure differences coupled with ionic transport in a charged matrix. We found that the simulated pathological architecture produced fragmented transport pathways and weaker electrical signals. This suggests that differences in tissue structure could influence simulated electrokinetic responses by altering connectivity and directional organization of the transport network.

Linking electrical measurements with the physical architecture of cartilage, our approach has advantages over purely mechanical descriptions. Mechanical stiffness alone does not uniquely reflect microstructural organization, whereas electrokinetic signals depend simultaneously on surface charge density, pore connectivity and fluid dynamics (Lu, Mow and Guo 2009; Kim et al. 2015; Zevenbergen et al. 2018; Collins et al. 2021; Sun et al. 2021; Żylińska et al. 2021). Thi suggests that electrical measurements may provide a more sensitive indicator of early structural damage.

Our approach differs from descriptions of cartilage electrokinetics relying on continuum representations of the extracellular matrix and employing homogenized material parameters. Instead of assigning averaged transport coefficients to the entire tissue, we derive a discrete network from the spatial distribution of histological features and use this geometry to compute fluid transport and electrical potentials. Our analysis includes node connectivity, directional transport pathways and spatially resolved electrokinetic potentials, which are not typically obtained in continuum models of cartilage mechanics.

Our study has limitations. The pore-network reconstruction from a single histology image does not represent the full pore architecture of cartilage, since most fluid pathways occur at nanometer scales that cannot be resolved by light microscopy. The network nodes correspond to lacunae rather than to actual microscopic pores in the extracellular matrix. Our statistical and electrical plots are illustrative models rather than analyses derived from experimental datasets. The mathematical relations used for fluid transport and electrokinetic potentials have not been validated against empirical measurements of cartilage electrokinetic signals. The contributions of nanoscale matrix channels and larger lacunar structures in generating electrokinetic potentials remain unclear. It is also uncertain how variations in proteoglycan concentration could modify the spatial propagation of electrical signals through the tissue.

Experimental previsions follow from our simulations. First, compression of intact articular cartilage should produce electrokinetic potentials whose magnitude increases with the applied mechanical strain. If pressure-driven ion transport governs signal generation, the measured electrical potential should scale with the imposed pressure gradient across the tissue thickness.

Second, we predict a spatial gradient of electrical signal intensity across cartilage depth. Microelectrode measurements inserted at controlled depths should detect systematic differences in potential magnitude between superficial, middle and deep layers, reflecting the heterogeneous organization of the extracellular matrix.

Third, degenerative structural alterations should modify both the amplitude and temporal evolution of the signal. When equivalent loads are applied to healthy and degenerated cartilage samples, the pathological tissue should exhibit lower peak potentials and faster signal decay due to increased permeability and disrupted matrix organization.

Still, electrokinetic potentials should scale with the microstructural connectivity of the extracellular matrix: samples with higher lacunar connectivity and smaller effective pore spacing are expected to generate larger streaming potentials under identical mechanical loads, because fluid transport pathways become more coherent.

We also predict directional anisotropy of electrical signals. Potentials measured under compression parallel to the dominant collagen orientation should differ systematically from those measured under orthogonal loading directions, reflecting anisotropic permeability and transport within the collagen network.

Furthermore, spatial clustering of lacunae should generate localized electrokinetic hotspots where multiple transport pathways converge, producing measurable heterogeneity in electrical potentials across the tissue surface.

Variations in matrix charge density introduce another measurable prediction. Proteoglycan depletion should alter the electrokinetic response in a nonlinear manner, with signal amplitudes decreasing more rapidly than predicted by simple proportional scaling with fixed charge density.

From a practical perspective, integrating histological reconstruction with electrokinetic modeling could enable the derivation of quantitative descriptors of cartilage architecture that are directly linked to measurable electrical behavior. Our approach could assist in interpreting experimental measurements obtained from electromechanical cartilage testing systems, by providing structural explanations for observed signal patterns. Structural descriptors derived from reconstructed networks might also facilitate comparison among samples with different degrees of tissue alteration. Computational representations derived from imaging could support systematic exploration of how matrix organization influences mechanical–electrical coupling in hydrated biological materials.

In conclusion, we investigated whether histological organization of cartilage can be translated into computational representations that reproduce electrokinetic responses generated during mechanical loading. Our simulations comparing physiological and degenerative structures indicate that variations in tissue architecture correspond to differences in predicted electrical behavior. Therefore, linking histological structure with transport modeling could provide a way to relate tissue organization with electrically measurable signals.

## DECLARATIONS

**Ethics approval and consent to participate.** This research does not contain any studies with human participants or animals performed by the Author.
**Consent for publication.** The Author transfers all copyright ownership in the event the work is published. The undersigned author warrants that the article is original, does not infringe on any copyright or other proprietary right of any third part, is not under consideration by another journal and has not been previously published.
**Availability of data and materials.** All data and materials generated or analyzed during this study are included in the manuscript. The Author had full access to all the data in the study and took responsibility for the integrity of the data and the accuracy of the data analysis.
**Competing interests.** The Author does not have any known or potential conflict of interest including any financial, personal or other relationships with other people or organizations within three years of beginning the submitted work that could inappropriately influence or be perceived to influence their work.
**Disclaimer**. The views expressed are those of the author and do not necessarily reflect those of the affiliated institutions.
**Funding.** This research did not receive any specific grant from funding agencies in the public, commercial or not-for-profit sectors.
**Acknowledgements:** none.
**Authors' contributions.** The Author performed: study concept and design, acquisition of data, analysis and interpretation of data, drafting of the manuscript, critical revision of the manuscript for important intellectual content, statistical analysis, obtained funding, administrative, technical and material support, study supervision.
**Declaration of generative AI and AI-assisted technologies in the writing process.** During the preparation of this work, the author used ChatGPT 5.2 to assist with data analysis and manuscript drafting and to improve spelling, grammar and general editing. After using this tool, the author reviewed and edited the content as needed, taking full responsibility for the content of the publication.

## REFERENCES

1) Chen, Annan, Ziqin Wang, Zhizi Guan, Jiajun Wu, Qi Wei Shi, Senlin Wang, Yusheng Shi, Bin Su, Chunze Yan, Zuankai Wang and Jian Lu. 2026. "Echinoderm Stereom Gradient Structures Enable Mechanoelectrical Perception." *Nature* 651: 371–376. https://doi.org/10.1038/s41586-026-10164-9
2) Collins, A. T., G. Hu, H. Newman, M. H. Reinsvold, M. R. Goldsmith, J. N. Twomey-Kozak, H. A. Leddy, D. Sharma, L. Shen, L. E. DeFrate and C. M. Karner. 2021. "Obesity Alters the Collagen Organization and Mechanical Properties of Murine Cartilage." *Scientific Reports* 11 (1): 1626. https://doi.org/10.1038/s41598-020-80599-1
3) Farooqi, Abdul Razzaq, Rainer Bader and Ursula van Rienen. 2019. "Numerical Study on Electromechanics in Cartilage Tissue with Respect to Its Electrical Properties." *Tissue Engineering Part B: Reviews* 25 (2): 152–166. https://doi.org/10.1089/ten.teb.2018.0214
4) Kim, J. H., A. Hamamoto, N. Kiyohara and B. J. Wong. 2015. "Model to Estimate Threshold Mechanical Stability of Lower Lateral Cartilage." *JAMA Facial Plastic Surgery* 17 (4): 245–250. https://doi.org/10.1001/jamafacial.2015.0255
5) Lee, Jae-Hyun, Ye-Seul Jang and Won-Du Chang. 2025. "The Cartilage-Generated Bioelectric Potentials Induced by Dynamic Joint Movement: An Exploratory Study." *BMC Musculoskeletal Disorders* 26: 669. https://doi.org/10.1186/s12891-025-08939-8

6) Liang, X., X. Wang, Q. Xu, Y. Lu, Y. Zhang, H. Xia, A. Lu and L. Zhang. 2018. "Rubbery Chitosan/Carrageenan Hydrogels Constructed through an Electroneutrality System and Their Potential Application as Cartilage Scaffolds." *Biomacromolecules* 19 (2): 340–352. https://doi.org/10.1021/acs.biomac.7b01456
7) Lu, X. L., V. C. Mow and X. E. Guo. 2009. "Proteoglycans and Mechanical Behavior of Condylar Cartilage." *Journal of Dental Research* 88 (3): 244–248. https://doi.org/10.1177/0022034508330432
8) Miguel, Filipe, Frederico Barbosa, Frederico Castelo Ferreira and João Carlos Silva. 2022. "Electrically Conductive Hydrogels for Articular Cartilage Tissue Engineering." *Gels* 8 (11): 710. https://doi.org/10.3390/gels8110710
9) Nerger, Bryan A., Kirti Kashyap, Brendan T. Deveney, et al. 2024. "Tuning Porosity of Macroporous Hydrogels Enables Rapid Rates of Stress Relaxation and Promotes Cell Expansion and Migration." *Proceedings of the National Academy of Sciences of the United States of America* 121 (45): e2410806121. https://doi.org/10.1073/pnas.2410806121
10) Offeddu, G. S., I. Mela, P. Jeggle, R. M. Henderson, S. K. Smoukov and M. L. Oyen. 2017. "Cartilage-Like Electrostatic Stiffening of Responsive Cryogel Scaffolds." *Scientific Reports* 7: 42948. https://doi.org/10.1038/srep42948
11) Reynaud, Boris and Thomas M. Quinn. 2006. "Tensorial Electrokinetics in Articular Cartilage." *Biophysical Journal* 91 (6): 2349–2355. https://doi.org/10.1529/biophysj.106.082263
12) Sauerwein, Malte and Holger Steeb. 2020. "Modeling of Dynamic Hydrogel Swelling within the Pore Space of a Porous Medium." *International Journal of Engineering Science* 155: 103353. https://doi.org/10.1016/j.ijengsci.2020.103353
13) Schmidt-Rohlfing, Bernhard, Ulrich Schneider, Hans Goost and Jiri Silny. 2002. "Mechanically Induced Electrical Potentials of Articular Cartilage." *Journal of Biomechanics* 35 (4): 475–482. https://doi.org/10.1016/S0021-9290(01)00232-9
14) Sun, H., J. Zhou, Q. Wang, H. Jiang and Q. Yang. 2021. "Contribution of Perichondrium to the Mechanical Properties of Auricular Cartilage." *Journal of Biomechanics* 126: 110638. https://doi.org/10.1016/j.jbiomech.2021.110638
15) Sun, Y., K. Zhang, H. Dong, et al. 2022. "Layered Mechanical and Electrical Properties of Porcine Articular Cartilage." *Medical & Biological Engineering & Computing* 60: 3019–3028. https://doi.org/10.1007/s11517-022-02653-6
16) Tanikawa, Satoshi, Yuki Ebisu, Tomáš Sedlačík, Shingo Semba, Takayuki Nonoyama, Takayuki Kurokawa, Akira Hirota, Taiga Takahashi, Kazushi Yamaguchi, Masamichi Imajo, Hinako Kato, Takuya Nishimura, Zen-ichi Tanei, Masumi Tsuda, Tomomi Nemoto, Jian Ping Gong and Shinya Tanaka. 2023. "Engineering of an Electrically Charged Hydrogel Implanted into a Traumatic Brain Injury Model for Stepwise Neuronal Tissue Reconstruction." *Scientific Reports* 13: 2233. https://doi.org/10.1038/s41598-023-28870-z
17) Walker, J. C., A. M. Jorgensen, A. Sarkar, S. P. Gent and M. A. Messerli. 2022. "Anionic Polymers Amplify Electrokinetic Perfusion through Extracellular Matrices." *Frontiers in Bioengineering and Biotechnology* 10: 983317. https://doi.org/10.3389/fbioe.2022.983317
18) Wiese, Monika, Theresa Lohaus, Jan Haussmann and Matthias Wessling. 2019. "Charged Microgels Adsorbed on Porous Membranes: A Study of Their Mobility and Molecular Retention." *Journal of Membrane Science* 588: 117190. https://doi.org/10.1016/j.memsci.2019.117190
19) Youn, J. I., T. Akkin and T. E. Milner. 2004. "Electrokinetic Measurement of Cartilage Using Differential Phase Optical Coherence Tomography." *Physiological Measurement* 25 (1): 85–95. https://doi.org/10.1088/0967-3334/25/1/008
20) Zevenbergen, L., C. R. Smith, S. Van Rossom, D. G. Thelen, N. Famaey, J. Vander Sloten and I. Jonkers. 2018. "Cartilage Defect Location and Stiffness Predispose the Tibiofemoral Joint to Aberrant Loading Conditions during Stance Phase of Gait." *PLoS One* 13 (10): e0205842. https://doi.org/10.1371/journal.pone.0205842
21) Żylińska, B., A. Sobczyńska-Rak, U. Lisiecka, E. Stodolak-Zych, Ł. Jarosz and T. Szponder. 2021. "Structure and Pathologies of Articular Cartilage." *In Vivo* 35 (3): 1355–1363. https://doi.org/10.21873/invivo.12388